\documentclass[letterpaper,twocolumn,10pt]{article}
\usepackage{usenix-2020-09}

\usepackage{cite}
\usepackage{amsmath,amssymb,amsfonts,url}
\usepackage{algorithmic}
\usepackage{graphicx}
\usepackage{textcomp}
\usepackage{xcolor}
\usepackage{booktabs} 
\usepackage{hyperref}
\usepackage[inline]{enumitem}

\usepackage[utf8]{inputenc}

\usepackage{listings}
\usepackage{xcolor}

\definecolor{codegreen}{rgb}{0,0.6,0}
\definecolor{codegray}{rgb}{0.5,0.5,0.5}
\definecolor{codepurple}{rgb}{0.58,0,0.82}
\definecolor{backcolour}{rgb}{0.95,0.95,0.92}

\lstdefinestyle{mystyle}{
    backgroundcolor=\color{backcolour},   
    commentstyle=\color{codegreen},
    keywordstyle=\color{magenta},
    numberstyle=\tiny\color{codegray},
    stringstyle=\color{codepurple},
    basicstyle=\ttfamily\footnotesize,
    breakatwhitespace=false,         
    breaklines=true,                 
    captionpos=b,                  
    floatplacement=t,  
    keepspaces=true,                 
    numbers=left,                    
    numbersep=5pt,                  
    showspaces=false,                
    showstringspaces=false,
    showtabs=false,                  
    tabsize=2
}
\begin{document}

\title{ZCSD: a Computational Storage Device over Zoned Namespaces (ZNS) SSDs}

\author{
{\rm Corne Lukken, Giulia Frascaria, and Animesh Trivedi}\\
Vrije Universiteit Amsterdam
} 

\maketitle

\begin{abstract}

The Big Data trend is putting strain on modern storage systems, which have to support high-performance 
I/O accesses for the large quantities of data. With the prevalent Von Neumann computing architecture, 
this data is constantly moved back and forth between the computing (i.e., CPU) and storage entities 
(DRAM, Non-Volatile Memory NVM storage). Hence, as the data volume grows, this constant data movement between 
the CPU and storage devices has emerged as a key performance bottleneck. To improve the situation, researchers 
have advocated to leverage computational storage devices (CSDs), which offer a programmable interface to 
run user-defined data processing operations close to the storage without excessive data movement, thus 
offering performance improvements. However, despite its potential, building CSD-aware applications remains 
a challenging task due to the lack of exploration and experimentations with the right API and abstraction. This is due 
to the limited accessibility to latest CSD/NVM devices, emerging device interfaces, and closed-source software 
internals of the devices. To remedy the situation, in this work we present an open-source CSD prototype 
over emerging NVMe Zoned Namespaces (ZNS) SSDs and an interface that can be used to explore application designs for  
CSD/NVM storage devices. In this paper we summarize the current state of the practice with CSD devices, make a case 
for designing a CSD prototype with the ZNS interface and eBPF (ZCSD), and present our initial findings. 
The prototype is available at \url{https://github.com/Dantali0n/qemu-csd}.   
\end{abstract}


\section{Introduction}
In the age of Big Data, every day we generate 2.5 Exabytes ($10^{18}$) of data, and by 2025 this rate will grow to 
480 Exabytes/day (a 192-fold increase) to generate 175 Zettabytes of data per year~\cite{2021-idc-175-zbytes}.
With the prevalent Von Neumann computing architecture, this vast amount of data is stored and processed in separate 
architectural entities, with storage devices such as hard-disk drives (HDDs) or solid state drives (SSDs) only passively 
storing data~\cite{2018-neumann-bottleneck}. When needed, data is moved between storage devices and DRAM to be processed 
by the CPU. However, while the data growth is exploding, the CPU performance improvements and link speeds 
(over which data movement happens, e.g., PCIe or Ethernet) are not projected to improve at the same rate over 
the next 5 years. Consequently, the excessive data movement with a system for CPU-driven data processing has become 
the key performance bottleneck for data-intensive workloads~\cite{2014-micro-ndp}.


One promising avenue to solve the challenges associated with excessive data movement is to push compute to storage, 
rather than pull data from storage to compute (i.e., the CPU). Today researchers are actively exploring the possibility 
of \textit{``Programmable Storage''} or \textit{``Computational Storage''} 
devices (CSD)~\cite{2013-sigmod-smartssd,2016-isca-biscuit,2016-vldb-yoursql,2017-micro-summarizer,2013-fast-active-flash,2017-vldb-caribou,2014-osdi-willow,2014-vldb-ibex,2013-sc-active-flash,2019-tos-registor,2019-atc-insider,2020-tos-newport-csd,2020-middleware-blockNDP,2013-msst-ssd-mr}.
The idea of programmable storage is not new, and has been explored previously with 
HDDs~\cite{1998-asplos-active-disks,1997-sigmetrics-secure-disks,1998-vldb-active-storage,1998-sigmod-idisk}, though 
the low disk bandwidths dominated the data processing costs.
With the rise of Non-Volatile Memory (NVM) storage the idea is being revisited as NVM SSDs offer significant 
device-internal bandwidths~\cite{do2019programmable}, and already have an element of programmability to run 
management code like Flash Translation Layer (FTL) for NAND flash chips. The elimination (or reduction) of 
data movement due to storage programmability offers energy, cost, and performance benefits.

\begin{flushleft}
\begin{table*}[]
\begin{tabular}{p{0.13\textwidth} p{0.2\textwidth} p{0.2\textwidth} l p{0.3\textwidth} }
\toprule
\textbf{} & \textbf{Hardware} & \textbf{Software API} & \textbf{Open} & \textbf{Comments}
\\
\midrule
Smart SSDs~\cite{2013-msst-ssd-mr} & RISC processor & MapReduce & No & usex an object-based communication protocol, OS/application transparent \\ 
Active Flash~\cite{2013-fast-active-flash} & ARM SATA controller & Statistical analysis functions & No & analysis functions fused with the FTL, not possible to change them \\ 
Intelligent SSDs~\cite{2013-sc-active-flash} & ASICs, reconfigurable stream processors & Map-Reduce, OpenCL & No & analytics, statistical and database scan and filter operators implemented \\
SmartSSD~\cite{2013-sigmod-smartssd} & Embedded SSD CPU & 3 new commands for a session-based protocol & No & Pre-compiled selection and aggregation operators into the SSD FTL \\
Willow~\cite{2014-osdi-willow} & RISC processor  & RPC / Client-server & No & Supports arbitrary user code and integration with the host file system\\
Ibex~\cite{2014-vldb-ibex} & FPGA & MySQL storage engine & No & FPGA as implicit co-processor, groupby and filter offload to an FPGA \\
Biscuit~\cite{2016-isca-biscuit} & ARM Cortex R7 cores & Dataflow operators & No & Supports C++11, I/O queues, multithreading, and dynamic memory in an SSD\\
Caribou~\cite{2017-vldb-caribou} (also~\cite{2016-vldb-bluecache}) & FPGA & Key-Value operations & Yes & focus on key-value store operations (get, put, delete, scan) only \\
YourSQL~\cite{2016-vldb-yoursql} & ARM Cortex R7 cores & Database query offloading & No & full database integration with MariaDB query plans for query offloading \\
Summarizer~\cite{2017-micro-summarizer} & ARM cores/FPGA & user functions on flash pages & No & supports opportunistic offloading of code to SSDs, no file system integration \\  
Registor~\cite{2019-tos-registor} & FPGA with RegEx IC modules & specific calls to process files & No & workload-specific acceleration (regex only) \\
INSIDER~\cite{2019-atc-insider} & FGPA & RPC/Client-server & Yes & Full integration with the host file system, virtual file abstraction, requires FPGA knowhow \\
CognitiveSSD~\cite{2019-atc-cognitive-ssd} & OpenSSD~\cite{2021-openssd-project} (FPGA with ARM Cortex-A9 cores) & specific r/w calls on flash PBAs & Yes & workload-specific (feature extraction, deep-learning, and graph search) offloading\\ 
NGD Newport~\cite{2020-tos-newport-csd} & ARM cortex A53 (64-bits) & RPC / Client-server & No & general-purpose, enterprise-grade, CSDs appear as a networked machine with attached storage \\
blockNDP~\cite{2020-middleware-blockNDP} & OpenSSD~\cite{2021-openssd-project} (FPGA with ARM Cortex-A9 cores) & eBPF code, shared memory & Not yet & supports general-purpose eBPF code execution on (read/write/transform) files inside a SSD \\
\midrule
ZCSD (\textit{work-in-progress}) & Any (eBPF-supported), QEMU host CPU & eBPF code, shared memory & Yes & Supports ZNS NVMe devices, integration with files/SPDK, uses NVMe command set\\
\bottomrule
\end{tabular}
\vspace{0.1cm}
\caption{Related work and comparison of recent CSD projects including work-in-progress ZCSD prototype (this work).}
\vspace{-0.5cm}
\label{tab:csd-compare}
\end{table*}
\end{flushleft}

Despite its potential, building applications for programmable storage remains challenging due to multiple 
reasons~\cite{barbalacecomputational}.  
First, there is no de-facto device design for CSD devices.Previous efforts have explored delivering programmability 
with the use of emdebbed CPUs~\cite{2014-osdi-willow}, FGPA~\cite{2019-atc-insider,2014-vldb-ibex}, 
ASICs/ARM cores~\cite{2020-tos-newport-csd,2013-sigmod-smartssd}, and there are on-going efforts to work on 
standardization from SNIA~\cite{2021-csd-snia}. A few closed-source CSD devices are available in the market 
now, albeit with limited availability~\cite{2021-csd-3,2021-csd-2,2021-csd-1}. Due to the complexity and limited 
accessibility to the device internals, it is challenging to explore the right 
computational hardware model (ISA, microarchitectural properties), and its host integration interface.
Second, there are no standard CSD programming abstractions, APIs, or programming models (dataflow, client-server, 
and shared memory)~\cite{barbalacecomputational}. Often the device manufacturers select the programming model, 
rather than letting developers choose the one that is the most suitable for their workload. Choosing an appropriate 
programming model necessitate an end-to-end visibility in the application design, host storage stack, and CSD 
internals due to the closed-nature of devices, and the semantic gap between these components.  
A CSD device has limited visibility into application-level data structures 
and file system data layouts. Likewise, the application is oblivious to CSD internal FTL and GC operations that 
can potentially change the data location. 
Lastly, due to the shared and multi-tenant nature of storage devices, CSD prototypes  
dedicated to one client or application~\cite{2013-sigmod-smartssd,2019-tos-registor} without security and privacy 
mechanisms are of limited usability. 
As a result, there is a need for a systematic evaluation of design choices for a data-intensive application (e.g., 
databases, analytics, AI/ML, searching and indexing) for programmable storage devices.

To facilitate such exploration, we present a Zoned Computation Storage Device (ZCSD) software 
prototype in QEMU~\cite{2005-atc-qemu}. Two key decisions shape the ZCSD design:  

\subsection{Why Zoned Namespaces (ZNS) Devices} 
The ZNS interface is the emerging standard (ratified by the NVM Express consortium as of June 2020) 
that allows host applications to have visibility and control over data placement and garbage collection 
operations in SSDs~\cite{2021-zns,2021-hotos-zns}.
It is a standardized, technical successor of Open-Channel~\cite{2017-fast-lightnvm}, 
streaming~\cite{2014-hotstorage-sssd}, and Software-defined SSDs~\cite{2014-asplos-sdf}.
The key features of the ZNS interface are (i) it does not allow in-place updates to 
written data; and (ii) zone reset and garbage cleaning must be done by the host software. 
As a result, the host software (including file systems) must become aware of \textit{append-only} properties 
of data stored on ZNS devices. 
Such append-only properties with visibility and control over SSD devices are very attractive to 
design a clear end-to-end APIs, and programming models with data consistency models, which was not possible 
before. We currently lack such end-to-end consistency models~\cite{barbalacecomputational}.
For example, multiple versions of data inside a log-structured file systems (e.g., F2FS) and ZNS can be trivially 
used to build a write-once data semantics model with shared memory.  



\subsection{Why eBPF for programmability}
With the notion of programmability we mean ability to run user-provided code in a dynamic and safe manner. 
Programmability can be supported by different software (kernel [kernel modules], file system [integration with the VFS], 
runtime and languages [Webassembely, Rust]) and hardware (FPGAs, ASICs, embedded CPUs) targets in the storage stack.
Diverse user requirements for open CSD exploration necessitate not being limited by restrictions from one particular 
``programmability'' implementation. 
Naturally, the closer to storage that we can run the code (Near-Data Processing) the better it is for performance. In 
this work, we advocate to use eBPF (extended Berkeley Packet Filter) language and toolchain~\cite{1993-atc-bpf,2021-ebpf-io} 
for three key reasons. 
First, eBPF instructed set is not tied to  
an application-domain like databases and it has already been used in networking~\cite{xdp}, tracing~\cite{bpf-tracing},
security~\cite{bpf-seccomp}, and even storage~\cite{bijlani2019extension,barbalace2019extos,2020-middleware-blockNDP,2020-arxiv-appcode,2021-arxiv-exo-bpf-storage}.
Second, due to the simplified nature of the eBPF instruction set, it is possible to verify for correctness and 
bounded execution of extensions. The Linux kernel already ships with an eBPF verifier~\cite{2018-kernel-bpf-verifier}, 
and multiple other prototypes are available~\cite{2019-pldi-bpf,ubpf}. Lastly, eBPF can support highly efficient 
code generation (via JITing) for multiple hardware devices such as x86, ARM, or FPGAs (the popular choices with CSDs),
and has been positioned as the unified ISA for heterogenous computing~\cite{2020-osdi-hxdp,bpf-uapi}. 
Hence, we choose to use eBPF to support programmability with ZNS SSDs.   

Our key contributions in this work are:
\begin{itemize}
  \item An analysis of current CSD state of the research and practices, and making a case for Zoned Computational Storage Devices (ZCSD);  
  \item A flexible open-source implementation of a ZCSD device in QEMU, \url{https://github.com/Dantali0n/qemu-csd}.
  \item An initial prototype evaluation for (i) performance of the prototype; and (ii) performance of the toolchain.    
\end{itemize}

\section{Current State-of-the-Art in CSD Research}
To the best of our knowledge, there are no open prototypes available for CSD at this moment that let users explore 
design choices without significant hardware or software development efforts (see Table~\ref{tab:csd-compare}). 
For example, FPGA related projects~\cite{2019-atc-insider,2014-vldb-ibex} require deep FGPA expertise to design and 
develop applications to run inside a CSD device. Many other CPU-driven prototypes are either not openly available, 
or are closely integrated with a specific-application domain like databases, thus, reducing the possibility for 
exploration. NGD Newport~\cite{2020-tos-newport-csd}, a enterprise-grade CSD device, offers a true general-purpose 
CSD design where the device run a full-fledged Linux environment on the CSD.
blockNDP is the closest work to ZCSD, which uses eBPF to offer general-purpose programmability to CSD, but as of 
this writing it is not yet open-source\footnote{\url{https://github.com/systems-nuts/blockNDP}, promised to be since \mbox{December} 2020. One part of the code (released as of Oct 17th, 2021) \textit{``developed outside Huawei are already available, but may contain bugs and insufficient instructions''}.}. 
Moreover,
none of the aforementioned prototypes offer support for new SSD interfaces such as Zoned Namespaces 
(ZNS) that can help to reduce the semantic gap between an application and the CSD device.

	\begin{figure} 	
	\centering
		\includegraphics[width=0.45\textwidth]{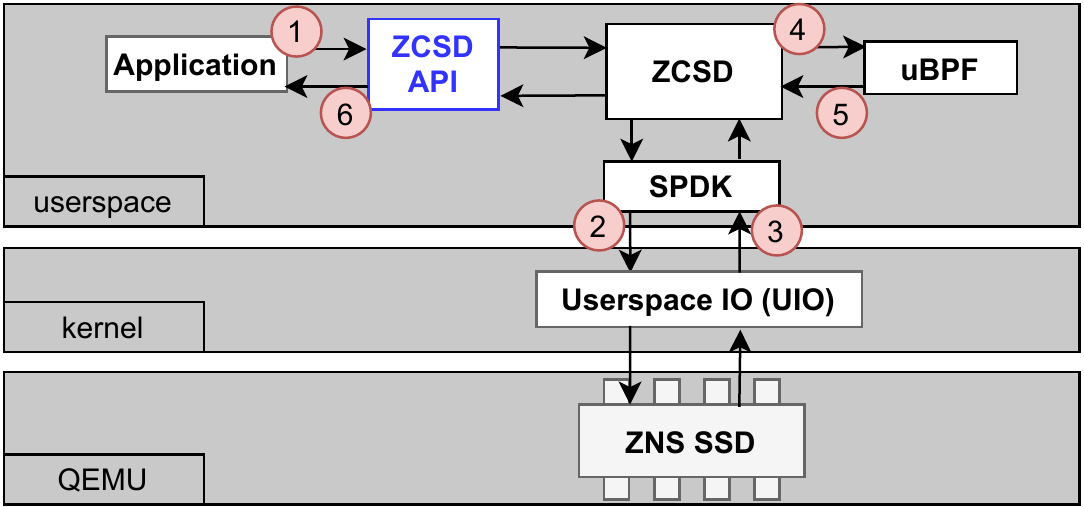}
		\caption{Overview of the components in the ZCSD prototype.}
		\label{fig:prototypesteps}
	\end{figure}

\section{The ZCSD Prototype}
In this section we present the design and implementation details of an open-sourced, Zoned Computation Storage 
Device (ZCSD). ZCSD is designed for 
(i) \textit{easy programmability} to present users a familar language, runtime, and toolchain;  
(ii) \textit{easy extensibility} of the prototype to allow users to extend and define new NVMe command sets, type of code
to run in CSD, and work split between the host application and ZCSD device;
(iii) \textit{easy analysis} to collect data about the execution of a prototype and reason about the design efficiency. 
ZCSD uses a NVMe Zoned SSD device in QEMU, and enhances it with an eBPF toolchain to support easy programmability and
extensibility. The whole prototype is written in C/C++, thus providing a familiar user language and toolchain. 
ZCSD uses an Userspace eBPF VM (uBPF)~\cite{ubpf}, which allows us to JIT and run general-purpose eBPF code
in ZCSD. An alternative would have been to choose the kernel eBPF infrastructure, but it is known to be restrictive 
in multiple ways such as restrictions on the number of instructions, absence of accesses to user data 
structures, and limited execution environment (access to function hooks, parameters and return codes)~\cite{2020-giulia-ebpf-talk}.
To support analysis, the whole prototype collects multiple performance statistics such as runtime, number of instructions
executed, JITing time, amount of data movement saved, etc. that can help user to reason about their design. 
\begin{figure}[!t]
\noindent\begin{minipage}[!t]{\linewidth}
\begin{lstlisting}[language=C++,caption={\label{tab:API} The two parts user-extensible ZCSD API.}]
namespace qemucsd::nvm_csd {
	class NvmCsd {
	public:
		NvmCsd(struct arguments::options*, struct spdk_init::ns_entry *);
		~NvmCsd();
		// part-i of the API between app and ZCSD 
		uint64_t nvm_cmd_bpf_run(void *bpf_elf, uint64_t bpf_elf_size);
		void nvm_cmd_bpf_result(void *data);
	protected:
	    // part-2 of the API to define the eBPF interface
		static void bpf_return_data(void *data, uint64_t size);
		static void bpf_read(uint64_t lba, uint64_t offset, uint16_t limit, void *data);
		static uint64_t bpf_get_lba_siza(void);
		static void bpf_get_mem_info(void **mem_ptr, uint64_t *mem_size);
	};
}
\end{lstlisting}
\end{minipage}
\end{figure}

\textit{The workflow life cycle:} Figure~\ref{fig:prototypesteps} shows the overall steps in the process. A CSD
application interacts with the ZCSD device using a user-defined ZCSD API (step 1, discussed in the next paragraph). 
The API allows the
application to identify the data and associated eBPF program that should execute on the data. ZCSD prototype then reads
the necessary data from the ZNS SSD in the userspace (step 2 and 3). Currently, we use SPDK/UIO to directly access the
device using the ZNS NVMe command set, but integration with other access mechanisms such as \texttt{libnvme} or
\texttt{libzbd} are also possible. After accessing the data, the user provided eBPF program is checked and JITed using 
uBPF (step 4 and 5) on the read data. The final result of executing the eBPF program on the data is returned to the
user in the step 6. An important note here is that, in its current incarnation without any hardware
offloading, ZCSD can not eliminate data movement from the storage device to the application. However, with the 
prototype a developer can measure the extent of the data reduction possible with and without ZCSD support, thus,
making it possible to reason about viability of using CSD in their application.      
 

\textit{ZCSD API:} The ZCSD API allows developers to define 
(part-i) the nature of interaction with ZCSD device and application; (part-ii) what should be 
offloaded in ZCSD using eBPF. The two parts of the APIs are defined in the \texttt{NvmCSD} class 
(can be extended by user by C++ inheritance) and shown in listing~\ref{tab:API}. 
The part-i of the API (ZSCD-app) allows a user to attach an eBPF program, by including a eBPF-program ELF 
header at the compile time, running it synchronously (line 7), and getting results (line 8). In the future we wish 
to extend this to allow asynchronous execution, with a dynamic runtime compilation support as done with the Linux/BCC 
framework~\cite{2021-bcc}. 
The eBPF API part (also defined and extensible by the user) allows the eBPF
program to read and process data from any offset from the ZNS SSD (lines 11-14). The result of eBPF processing is 
returned to the user by calling \texttt{bpf\_return\_data}. The current prototype is focused on defining 
these extensible hooks that can be used by developers to control the nature of interaction with a ZCSD device.

\begin{center}
    \begin{figure}
    \centering
        \includegraphics[width=0.45\textwidth]{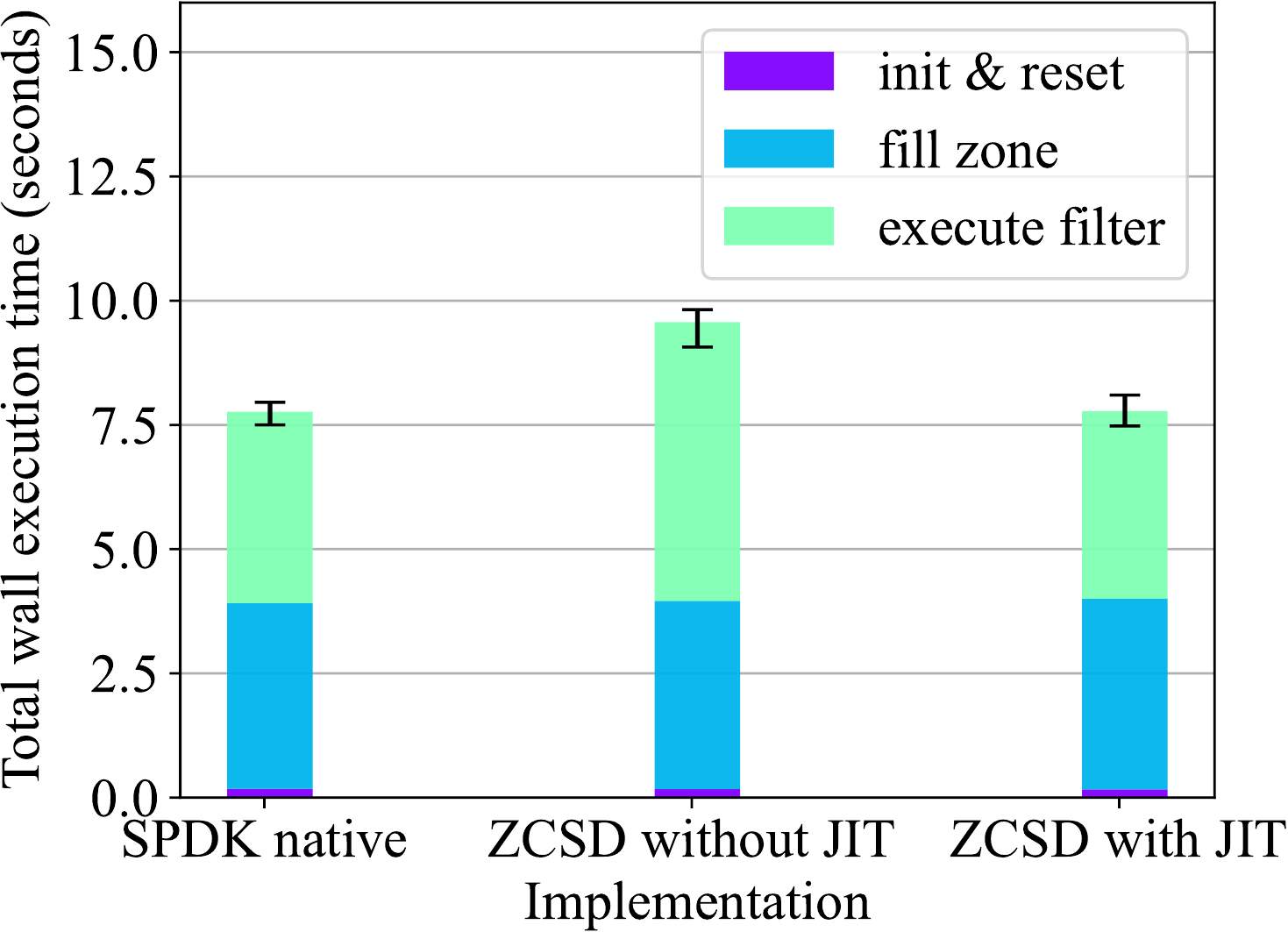}
        \caption{Overview of the performance across the different scenarios.}
        \label{fig:performance}
    \end{figure}
\end{center}

\section{Evaluation}
In this section we present an initial benchmarking about the performance and 
associated toolchain overheads of the ZCSD prototype with integer filter offloading.
The performance is evaluated across three different scenarios with the ZNS SSD device in QEMU:
\begin{enumerate*}
    \item SPDK without any computational capabilities, representing the current way in which applications can read and process data in userspace;
    \item ZCSD-uBPF without JIT, representing ZCSD device with a stack machine executing a single instruction at a time~\cite{ubpf-stack-machine};
    \item ZCSD-uBPF with JIT, representing an end-to-end compiled framework with (potential) JIT overheads. 
\end{enumerate*}
Scenarios 2 and 3 are different because uBPF performs memory bounds checking in the first case but not when executing JITed code. 

In all the testing scenarios we perform the same steps, namely initializing and resetting
the ZNS SSD, reading data, filtering, and returning the results to the user. The zone size is 256MB and the block/page 
size is 4kB (default values). The workload consists of filling the first zone with random integers ( 256MB/4 = 64 million 
integers) and counting the integers in the zone above $RAND\_MAX / 2$, thus probabilistically filtering half of 
the integers. The methods utilized to perform these operations are kept as uniform as possible to remove any 
measurement error and only measure the cost of the prototype and uBPF. The data is processed in the page granularity. 
An alternate architecture could have been reading the whole zone once for processing, however, we opted for a conservative design choice 
based on the page size in order to support limited memory resources of CSD devices which may not be able to cache large working 
dataset for processing. However, the access granularity is not fundamental, and we expect to extend the support with future work (discussed in the next section)

The primary goal of our current evaluation is to demonstrate that the footprint of the prototype is low and that it 
can yield performance gains with low software overhead.
Figure~\ref{fig:performance} shows our results. The x-axis shows the three benchmarked scenarios and the y-axis shows
the breakdown of the elapsed wall clock time (lower is better). The reported numbers are the averages of 5 runs, with
the whiskers showing the min and max runtimes. The key takeaway message from the graph is that our ZCSD prototype
performs comparable to a native SPDK implementation in the JITed mode. As expected, the time for initialization and
filling is the same in all scenarios as only the filtering is offloaded to uBPF. The execution of the filter code is
different, with the ZCSD implementation without JIT (configuration \#2) being the slowest due to its stack machine
implementation and memory-bound checking. When used with JIT (takes 152 $\mu$seconds) the performance is within 1\% of the SPDK performance.
Nonetheless we are aware that a more complex application with checks and verification will add overheads to our 
toolchain.

\section{Conclusion and On-Going Work}
In this paper we made a case and presented an open-sourced computational storage device prototype with zoned namespace 
SSDs and eBPF language. Our evaluation on the initial prototype, ZCSD, demonstrates minimum overheads from the toolchain 
and implementation.   
While the prototype is functional and allows to experiment with CSD capabilities, we are working to expand its 
capabilities to:
(i) integrate lightweight NVMe device access using \texttt{libnvme};
(ii) integrate ZCSD with a zoned-enabled file system for Flash storage like F2FS \cite{lee2015f2fs} to design a unified end-to-end data consistency model between an application, the file system, and a CSD;   
(iii) support for additional hardware backends such as ARM and FPGA, with a fast and efficient eBPF code verifiers~\cite{2019-pldi-bpf}; 
(iv) offer built-in support for popular application-level data structures like B+-trees, hash tables, and LSM trees for data processing inside ZCSD. 
These efforts will further improve the support for research experiments and more complex data 
processing workloads on the programmable storage device. ZCSD prototype is open-sourced and publicly available 
at \url{https://github.com/Dantali0n/qemu-csd}.


\bibliographystyle{plain}
\bibliography{main}

\vspace{12pt}

\end{document}